# X-ray magnetic circular dichroism spectra on the superparamagnetic transition metal-ion clusters $[Mn_{12}O_{12}(CH_3COO)_{16}(H_2O)_{24}].2CH_3COOH.4H_2O$ (Mn12) and $[Fe_8O_2(OH)_{12}(tacn)_6]Br_8\ 9H_2O$ (Fe8)

[1]Paolo Ghigna, [2]Andrea Campana, [2]Alessandro Lascialfari, [3]Andrea Caneschi, [3]Dante Gatteschi, [4]Alberto Tagliaferri, [4]Nicholas Brookes

[1]Dipartimento di Chimica Fisica, Università di Pavia, V.le Taramelli 16, I-27100, Pavia, Italy

[2]Dipartimento di Fisica "A. Volta", Università di Pavia, Via Bassi 4, I-27100, Pavia, Italy

[3]Dipartimento di Chimica, Università di Firenze, Via Maragliano 75, I-50144, Florence, Italy

[4]ESRF, B. P. 220, F-38043 Grenoble Cedex, France

## Abstract

The first X-ray magnetic circular dichroism (XMCD) spectra on the title compounds are presented. In Mn12 the Mn(III) and Mn(IV) contributions are clearly apparent and evidence the opposite spin arrangement of the two ions. For Fe8 the typical two peak structure of Fe(III) compounds is found. Quantitative analysis yielded for both compounds a negligible $<L_z>$. This is the first direct experimental evidence of the quenching of the angular momentum in these systems. For Fe8 we found $<S_z>$=1.14 in $\mu_B$ units per Fe atom, in reasonable agreement with the $S$=10 ground state assumed for in this system.

The investigation of molecular clusters has provided important information on quantum tunneling of the magnetization in nanosize objects [1]. The most relevant examples have been a cluster comprising eight manganese(III) and four manganese(IV) ions, Mn12 [2], and another one with eight iron ions, Fe8 [3]. Both have an $S$=10 ground state which is largely split at zero applied magnetic field to leave the $M_S=\pm10$ components lying lowest. Thermally assisted quantum tunneling was observed in both clusters and in Fe8 clear evidence of pure quantum tunneling of the magnetization was observed below 300 mK.

Despite many different measurements have been performed on both compounds still some problems remain open. Since in Mn12 manganese(III) ions are present, which in octahedral symmetry have a ground $^5E_g$ state, orbital contributions to the magnetic moment cannot be excluded. Further it would be desirable to have a direct determination of the moments of the individual ions, in order to better know the nature of the wavefunction of the magnetic ground state, an important piece of information for modelling the mechanism of quantum tunneling.

Amongst the other techniques for characterizing the electronic structure and magnetic interactions, L-edge X-ray absorption spectroscopy is increasingly attracting interest [4]: the $2p\rightarrow3d$ transitions at the L edges are dipole allowed and strong, especially if compared to the weak dipole forbidden (quadrupole allowed) $2p\rightarrow3d$ transitions characteristic of the K edges. In addition L-edges have smaller natural line widths and the possibility of a strong magnetic circular dichroism effect [5]. In this work the first X-ray magnetic circular dichroism (XMCD) spectra of the molecular superparamagnets Mn12 and Fe8, at the Mn-$L_{II,III}$ and Fe-$L_{II,III}$, respectively, are presented. The main interest of such a kind of measurements is that, with the help of quite simple summation rules [6, 7], they allow the separation of the orbital and spin parts of the magnetic moment of the ground state. In addition, the $2p$ core hole formed in final state of a $L_{II,III}$ edge X-ray absorption spectra interacts via Coulomb and exchange interactions with the $3d$ valence electrons, thus making this technique sensitive to the oxidation state. Thus, in principle, the contributions of Mn(IV) and Mn(III) to the magnetic moment of Mn12 can be separated. By looking at the magnetic structure of Mn12, it is easy to see that this separation gave us direct informations on the sign of the exchange coupling between Mn(III) and Mn(IV).

Mn12 and Fe8 were synthesized according to the literature method [2, 3], and identified by elemental analysis. Crystals of Mn12 were obtained by vapor diffusion of acetone in the reagent solution, then filtered and washed with acetone. Crystals of Fe8 were isolated by filtration after two weeks of slow evaporation of the solution. The XMCD spectra have been collected at the ID12B Dragon beamline at the European Synchrotron Radiation Facility (ESRF, Grenoble, France), in the total electron yield mode, which probes nearly a 100 Å layer at the energy of the Mn and Fe $L_{II,III}$ edges. The degree of circular polarization can be estimated to be near 85 % [8]. For the experiment the samples have been mixed with an almost equivalent amount (by volume) of graphite, ground in an agate mortar and then pressed to pellets. Each pellet have been then fixed to an aluminum sample holder by means of tantalum coils to ensure a proper electrical contact, and then transferred to the cold finger of a helium cryo-magnet, which was kept under ultra high vacuum ($10^{-9}$ torr). The energy resolution was set to 0.2 eV by adjusting the entrance slits. The measurements have been taken in the 4-100 K temperature range at different applied magnetic fields in the range 0.5-7 T. A correction was applied for incomplete polarization. The measurements demonstrated quite difficult due to sample degradation in the beam, the more damaged material being Mn12. To minimize the effects of sample degradation we moved the sample with respect to the beam after a proper amount of dichroism scans, and periodically changed the sample; in addition, between each scan at fixed *T* and *H*, a valve was closed to avoid sample damaging during the change of the experimental conditions. The lifetime of the sample in the beam can be estimated to be 32 and 52 dichroism scans for Mn12 and Fe8, respectively.

Fig. 1 shows the Mn-$L_{II,III}$ edge spectra of Mn12 taken with two opposite photon angular momenta, at 5 K and in a field of 5 T. For comparison, in Fig. 2 the Mn-$L_{II,III}$ edge absorption spectra of $Mn_2O_3$ and $MnO_2$ are shown. These two spectra display the well known chemical shift of the absorption edges, according to which the spectral features are shifted to higher energy at higher oxidation states; this can be explained in terms of the screening by the valence electrons of the Coulomb interaction between the 2*p* initial states and the nucleus. The XMCD spectrum, obtained by the difference between the two spectra of Fig. 1 and normalizing at unit intensity of the peak at *c.a.* 642 eV, is shown in Fig. 3. By comparison with the spectra of the manganese oxides, the peaks in the dichroism between 640 and 642.5 eV are attributed to Mn(III), while the peak at c.a. 643.5 eV is attributed to Mn(IV). The opposite direction of the peaks attributed to Mn(III) and Mn(IV) confirms the antiferromagnetic coupling between the two subsets of ions in the Mn12 molecule. The opposite direction of the peaks was observed in the whole 4-100 K temperature range, a result which can be considered as unique. As an example in Fig. 4 the normalized XMCD intensities for the peaks at 642 eV and 643.5 eV are shown as a function of *H* and at different temperatures.

According to Thole *et al.* [6], a sum rule states that the integral of the XMCD signal over a complete core level edge is directly proportional to the expectation value of the orbital angular momentum in the ground state $\langle L_z \rangle$. In the case of spin orbit split edges like the $L_{II,III}$ the integral has to be taken over the two components: in the presence of an external magnetic field strong enough to magnetically saturate the sample, if *A* and *B* are the integrals of the dichroic signal under the $L_{III}$ and $L_{II}$ edges, respectively, $\langle L_z \rangle = -(2\mu_B/3C)\times[A+B]$, where *C* can be evaluated from the integral of the non-polarized spectrum. The integral curve of the XMCD spectrum of Fig. 3 is shown in the same figure as a dotted line. It is clearly apparent that the integral vanishes at the end of the spectrum. This gives the first direct experimental confirmation of the quenching of the angular momentum by the crystal field in the Mn12 molecule.

The Fe-$L_{II,III}$ edge spectra of Fe8 taken with two opposite photon angular momenta are shown in Fig. 5, for a field of 7 T and at $T$=4 K. The absorption spectra show the characteristic two peak structure which is typical of Fe(III) in an octahedral environment [9-11]. The corresponding dichroism, obtained from the difference of the two spectra of Fig. 5 and after normalization at unit intensity of the peak at c.a. 710 eV, is shown in Fig. 6. The integral curve of the dichroic signal is also shown in Fig. 6 as a dotted line, and also in this case, the integral becomes vanishingly small at the end of the spectrum, thus confirming also for Fe8 the quenching of the angular momentum by the crystal field. The trend with temperature and field of the intensity of the peak at *c.a.* 710 eV in the dichroism is shown in Fig. 7.

An other useful sum rule allows to extract the ground state expectation value of the spin momentum $<S_z>$ from the difference between the integrals of the dichroism over two spin orbit split edges [7]: $<S_z>=-(\mu_B/C)\times[A-2B]$, neglecting the intra-atomic magnetic dipole moment. In the case of the Mn-$L_{II,III}$ edges, the spin orbit coupling is too weak, and the $L_{II}$ and the $L_{III}$ states are quite mixed, and the spin momentum cannot be extracted reliably from the dichroic signal. Indeed, this can be done quite reasonably for Fe, for which the spin orbit coupling is sufficiently large. At $T$=4 K and $H$=7 T, for Fe8 we found $<S_z>=1.14$ in $\mu_B$ units per Fe atom, which would yield a total spin momentum of 9.12 in $\mu_B$ units. Considering that the system is not fully saturated at the minimum temperature (4 K) and that the maximum field that can be obtained now on ID-12B is 7 T, this value can be regarded to be in excellent agreement with the $S_z$=10 ground state commonly assumed for in this system [12].

In summary, we have presented the first X-ray magnetic circular dichroism (XMCD) spectra on the molecular superparamagnets $[Mn_{12}O_{12}(CH_3COO)_{16}(H_2O)_{24}]\cdot 2CH_3COOH\cdot 4H_2O$ (Mn12) and $[Fe_8O_2(OH)_{12}(tacn)_6]Br_8\ 9H_2O$ (Fe8), at the Mn-$L_{II,III}$ and at the Fe-$L_{II,III}$ edges, respectively. The main result of this investigation is the direct experimental confirmation of the quenching of the orbital momentum by the crystal field in both the compounds. Moreover, in Mn12 the opposite directions of the Mn(III) and Mn(IV) were clearly observed, thus confirming the proposed ground state spin structure. These opposite directions could be detected in the whole temperature range 4-100 K at each constant field. For Fe8 we have extracted from the dichroism signal a value of the expectation value for the spin moment in the ground state which is in good agreement with the $S$=10 in the ground state assumed for in this system.

# Figure captions

**Fig. 1. –** Mn-$L_{II,III}$ edge XAS spectra of Mn12 at 5 K and in a field of 5 T; the dotted and full lines correspond to the two different photon helicities.

**Fig. 2. –** Isotropic Mn-$L_{II,III}$ edge XAS spectra of $Mn_2O_3$ (full line) and $MnO_2$ (dotted line) at room temperature.

**Fig. 3. –** Mn-$L_{II,III}$ edge XMCD spectrum of Mn12 at 5 K and in a field of 5 T (full line), after normalization to 100 % polarization rate and unit intensity of the peak in the isotropic XAS spectrum at *c.a.* 642 eV. The dotted line is the running integral of the XMCD signal.

**Fig. 4. –**Normalized XMCD intensities for the peaks at 642 eV (circles) and 643.5 eV (squares), as a function of $H$ at the indicated $T$.

**Fig. 5. –** Fe-$L_{II,III}$ edge XAS spectra of Fe8 at 4 K and in a field of 7 T; the dotted and full lines correspond to the two different photon helicities.

**Fig. 6. –** Fe-$L_{II,III}$ edge XMCD spectrum of Fe8 at 4 K and in a field of 7 T (full line), after normalization to 100 % polarization rate and unit intensity of the peak in the isotropic XAS spectrum at *c.a.* 710 eV. The dotted line is the running integral of the XMCD signal.

**Fig. 7. –** Normalized XMCD intensity for the peak at 710 eV in the Fe8 compound at the indicated $H$ and as a function of $T$.

Figure 1

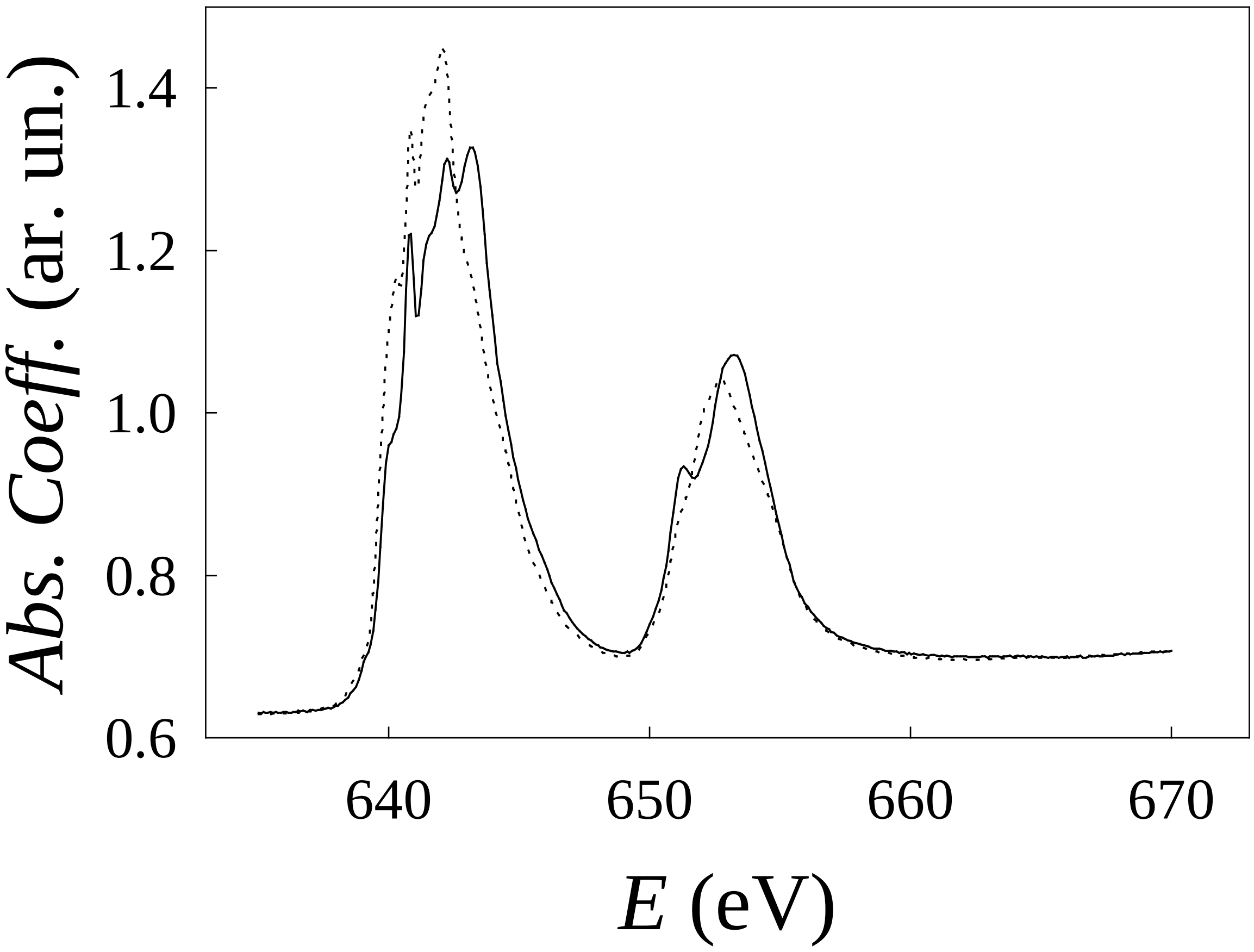

Figure 2

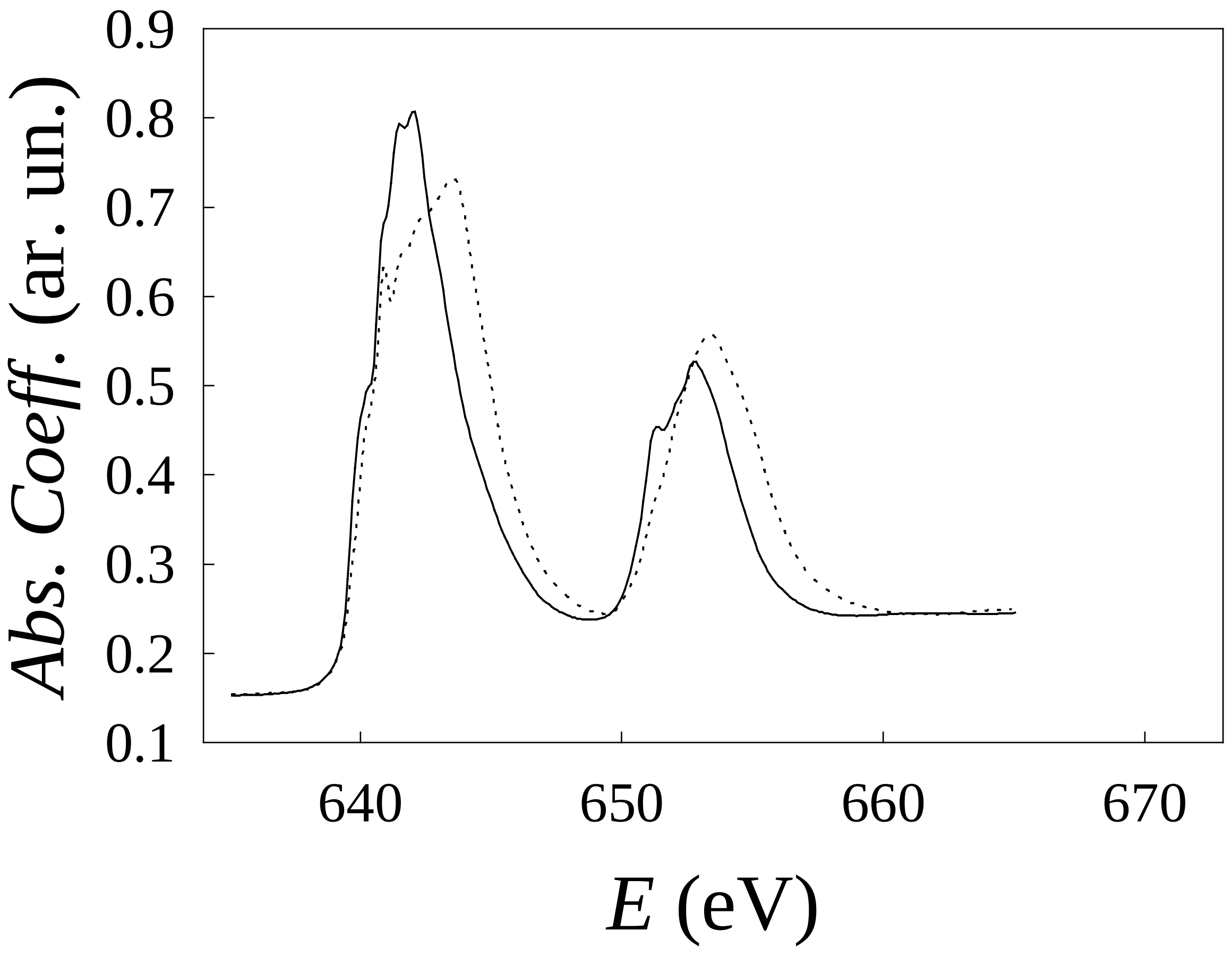

Figure 3

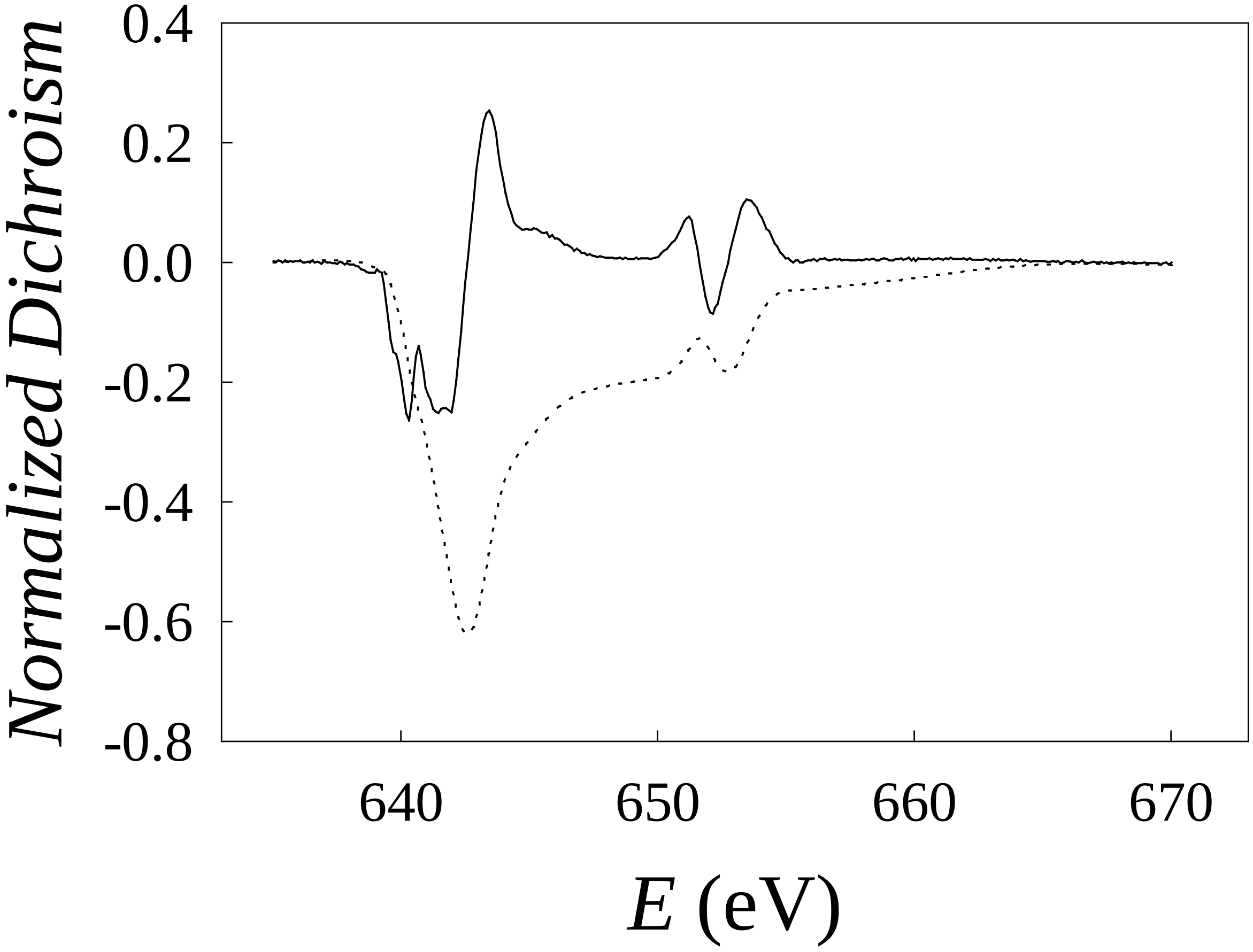

Figure 4

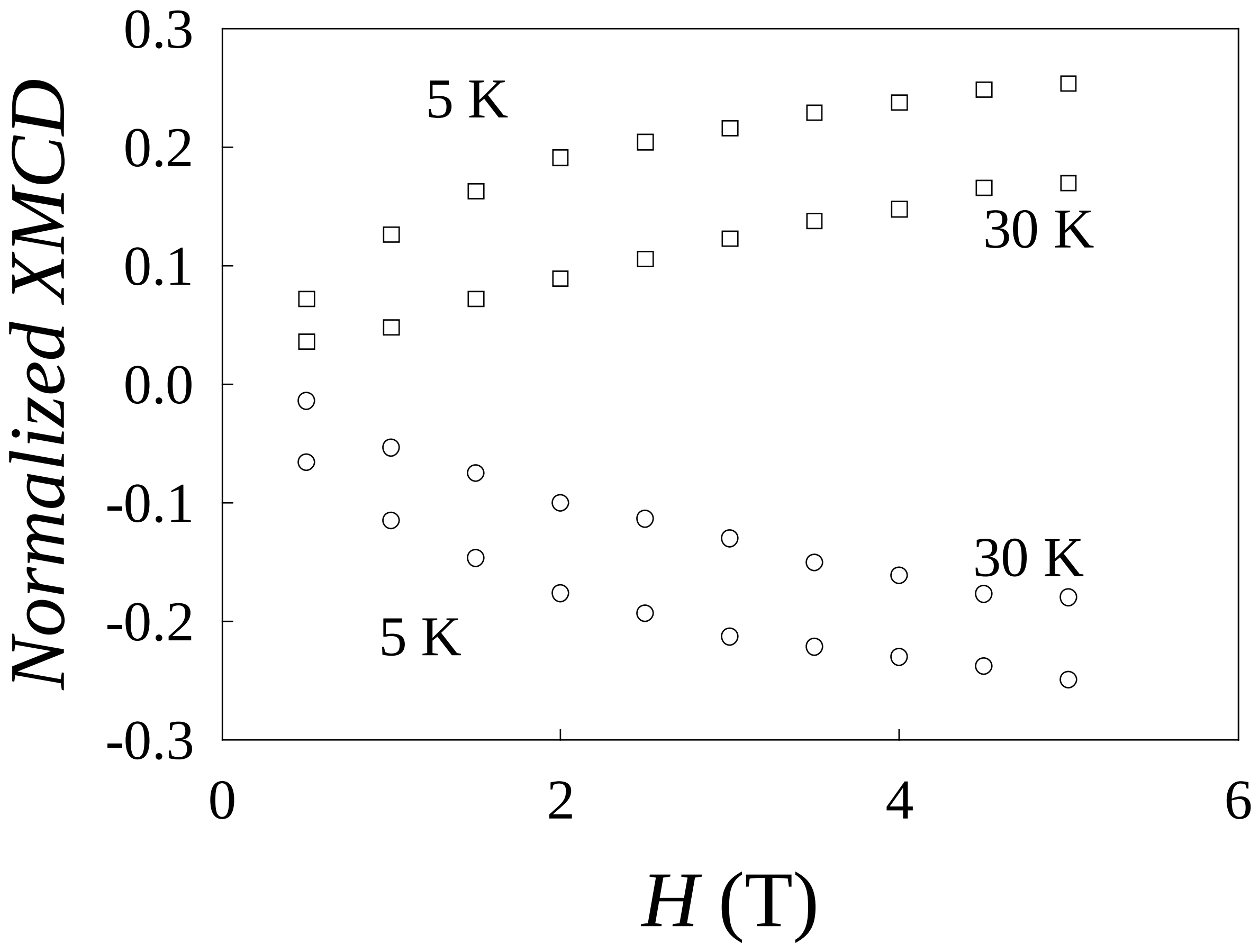

Figure 5

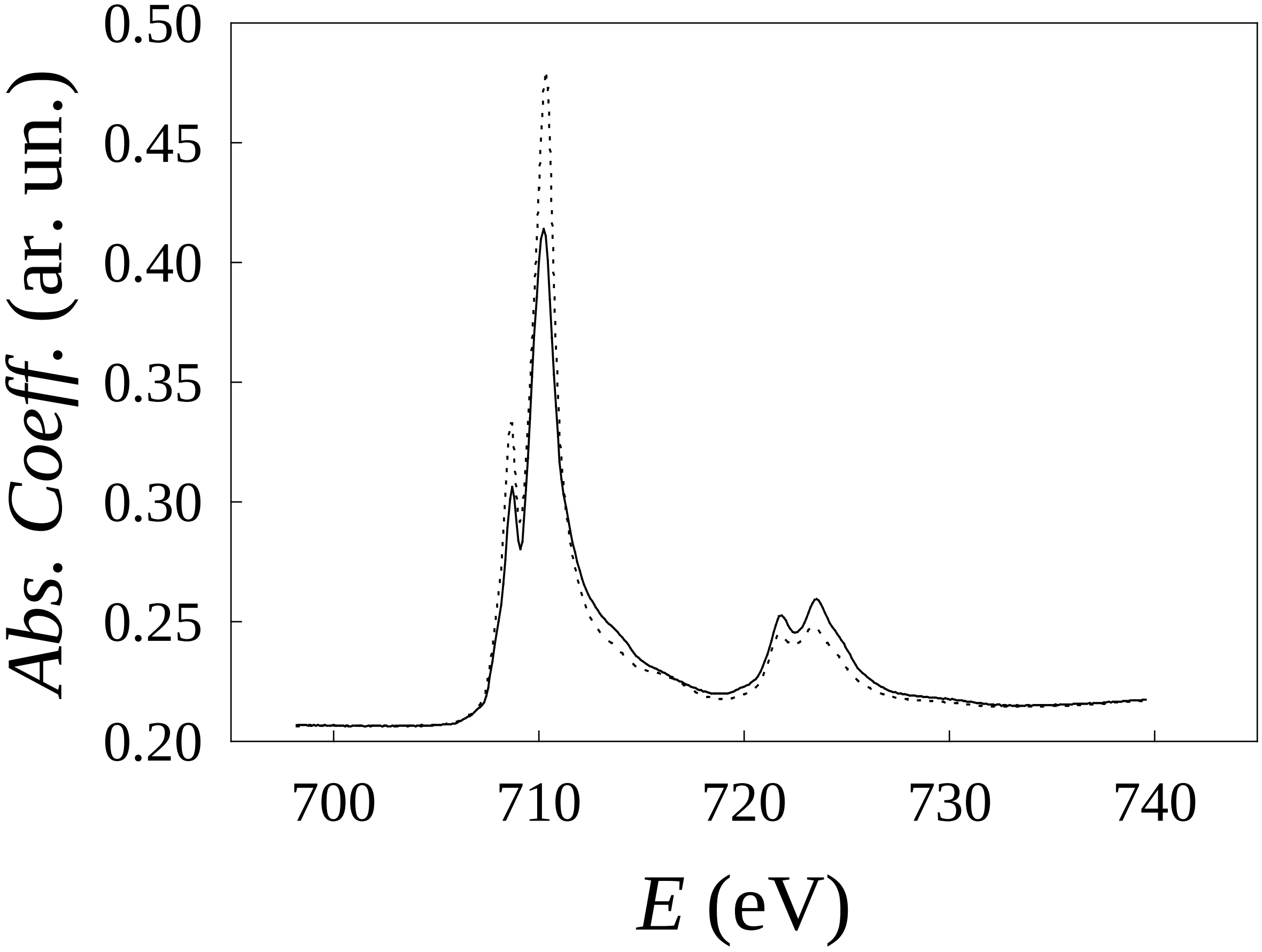

Figure 6

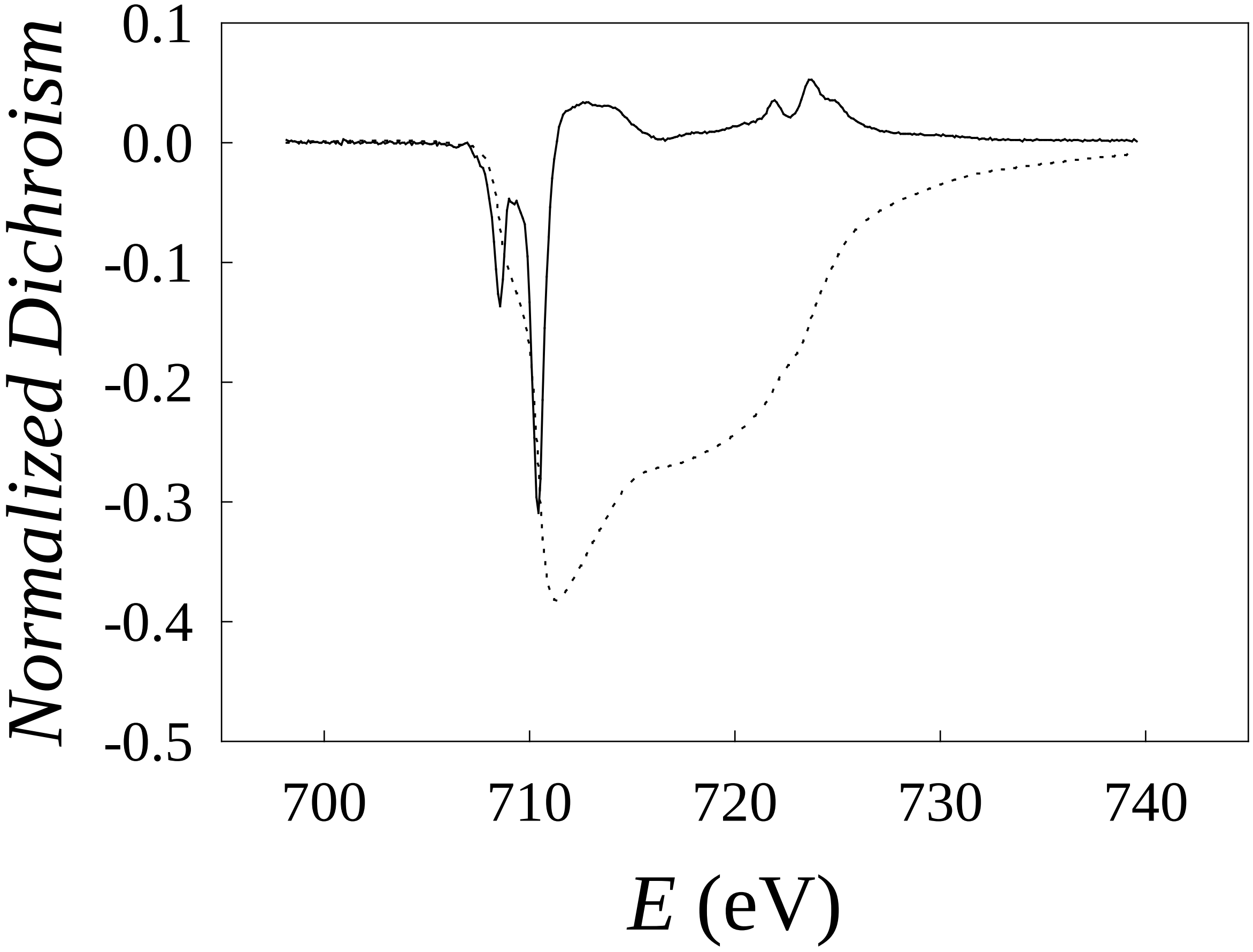

Figure 7

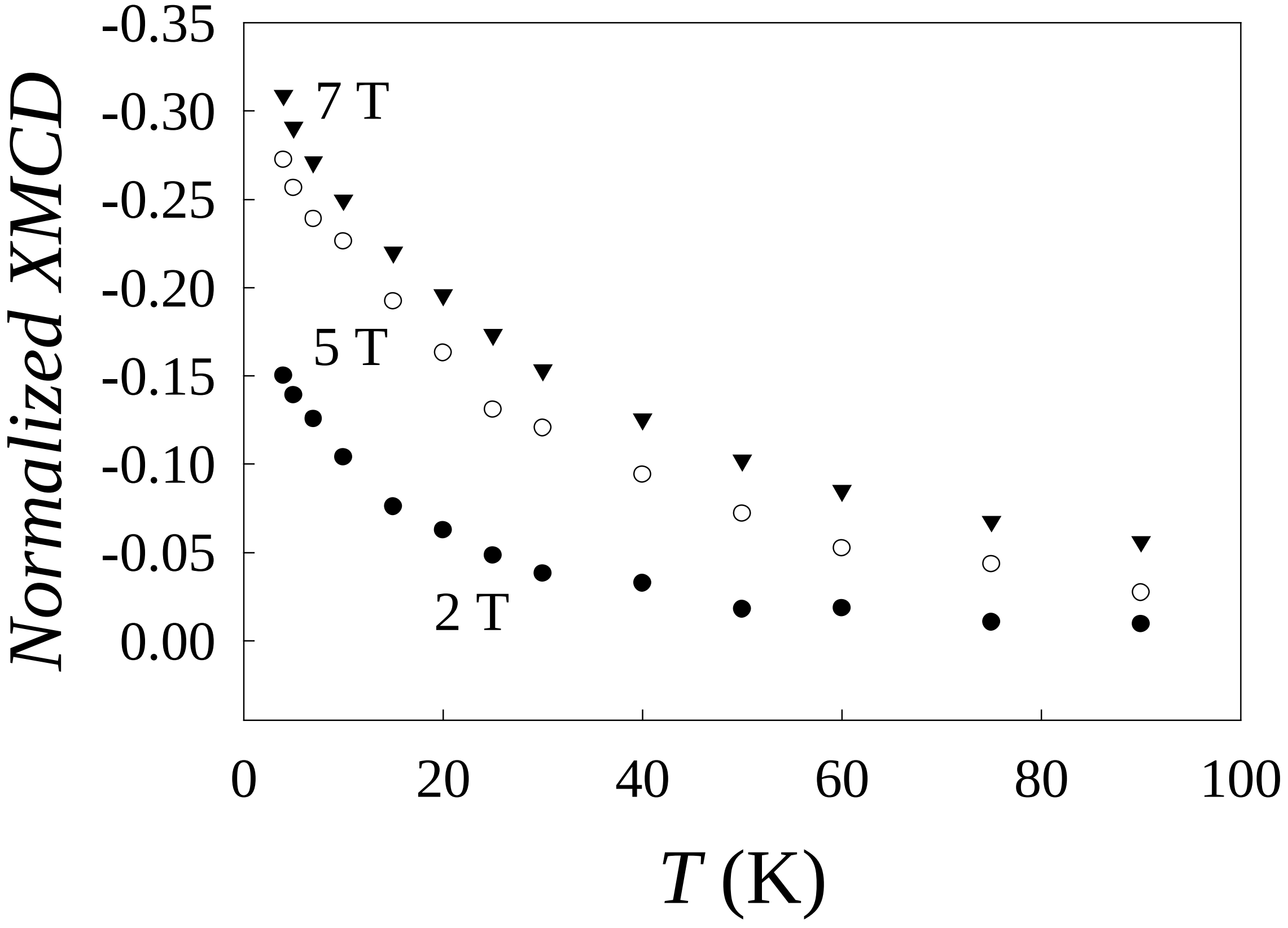